\documentclass{easychair}

\usepackage{graphicx,layout,amsmath,amssymb,bm}
\usepackage{amssymb}
\usepackage{amstext}
\newtheorem{algo}{Algorithm}
{\bfseries}{\itshape}
{\bfseries}{\itshape}
\newtheorem{theorem}{Theorem}{\bfseries}{\itshape}
\begin{document}

%
\title{Efficient and Low-Cost RFID Authentication Schemes}

%
\titlerunning{Efficient and Low-Cost RFID Authentication Schemes}


%
\author{Atsuko Miyaji\\
School of Information Science\\
Japan Advanced Institute of Science and Technology\\
1-1 Asahidai, Nomi, Ishikawa, Japan\\
\url{miyaji@jaist.ac.jp}\\
\and
Mohammad Shahriar Rahman\\
School of Information Science\\
Japan Advanced Institute of Science and Technology\\
1-1 Asahidai, Nomi, Ishikawa, Japan\\
\url{mohammad@jaist.ac.jp}\\
\and
Masakazu Soshi\\
School of Information Sciences\\
Hiroshima City University\\
3-4-1 Ozuka-Higashi, Asa-Minami-Ku, Hiroshima, Japan\\
\url{soshi@hiroshima-cu.ac.jp}\\
}

%
\authorrunning{A. Miyaji, M. S. Rahman, and M. Soshi}

\maketitle
\begin{abstract}
Security in passive resource-constrained Radio Frequency Identification (RFID) tags is of much interest nowadays. Supply-chain, 
inventory management are the areas where low-cost and secure batch-mode authentication of RFID tags is required. Resistance against illegal tracking, cloning, timing, and replay attacks are necessary for a secure RFID authentication scheme. Reader authentication is also necessary to thwart any illegal attempt to read the tags. With an objective to design a tracking, cloning, and replay attack resistant low-cost RFID authentication protocol, Gene Tsudik proposed a timestamp-based protocol using symmetric keys, named YA-TRAP$^*$. However, resistance against timing attack is very important for timestamp-based schemes, and the timestamps should be renewed in regular intervals to keep the tags operative. Although YA-TRAP$^*$ achieves its target security properties, it is susceptible to timing attacks, where 
the timestamp to be sent by the reader to the tag can be freely selected by an adversary. Moreover, in YA-TRAP$^*$, reader authentication is not provided, and a tag can become inoperative after exceeding its pre-stored threshold timestamp value. In this paper, we propose two mutual RFID authentication protocols that aim to improve YA-TRAP$^*$ by preventing timing attack, and by providing reader authentication. Also, a tag is allowed to refresh its pre-stored threshold value in our protocols, so that it does not become inoperative after exceeding the threshold. Our protocols also achieve other security properties like forward security, resistance against cloning, replay, and tracking attacks.
Moreover, the computation and communication costs are kept as low as possible for the tags. It is important to keep the communication cost as low as possible when many tags are authenticated in batch-mode. By introducing aggregate function for the reader-to-server communication, the communication cost is reduced. We also discuss different possible applications of our protocols. Our protocols thus capture more security properties and more efficiency than YA-TRAP$^*$. Finally, we show that our protocols can be implemented using the current standard low-cost RFID infrastructures.  
\end{abstract}
\maketitle

\section{Introduction}
\footnotetext[1]{A preliminary version of this paper appears at The 3rd International Workshop on Intelligent, Mobile and Internet Services in Ubiquitous Computing (IMIS 2009), IEEE. This is the full version.}

Radio-Frequency IDentification (RFID) is an automatic
identification method, relying on storing and remotely retrieving
data using devices called RFID tags or transponders. An
RFID tag is an object that can be applied to or incorporated
into a product, animal, or person, for the purpose of identification
using radio waves. Some tags can be read from
several meters away and beyond the line of sight of the reader.
RFID tags have opened the door to previously unexplored
applications. For example, in supply chains as suggested by EPC Global Inc. [\cite{EPC, juels}], baggage handling 
system at airports (e.g.,
the Hong Kong Airport), to locate people in amusement parks,
to combat the counterfeiting of expensive items \cite{JuelsPappu}, to trace 
livestock, to label books in libraries \cite{MolnerWagner} etc.
The use of RFID promises more flexible and intelligent handling of
consumer goods and devices. In an Intelligent Transport System (ITS), RFID tags are appropriate for electronic toll 
collection, airport access, airport ground transportation management, supply chain, vehicle-based technology solutions 
for parking, and security access control applications. The imminent ubiquity
of RFID tags, however, also poses a potentially widespread
threat to consumer privacy. 

In many RFID applications, it is necessary to read and
authenticate a large number of tags within a short period of
time. If RFID tags are easily readable, then tagged items will be subject to indiscriminate physical
tracking, as will their owners and bearers. A key to a safe and secure supply chain is the emphasis on
authenticating objects as well as tracking them efficiently
[\cite{AutoID} chapter 12]. However, unauthorized tracking of RFID tags is
viewed as a major privacy threat. In other words, we want tags to
reveal their identity to authorized RFID readers, so that the items can be tracked. However, for privacy, the
tags must not disclose their identity until the reader has been
authenticated; thus, the reader must authenticate itself to
the tags before doing anything else.

Cloned fake RFID tags and
malicious RFID readers pose major threats to the RFID-based system. The data on a genuine tag
can be easily scanned and copied by malicious RFID readers, and the copied data can be embedded into a fake tag. These
cloned fake tags can be attached to counterfeit products, which
can be introduced into a genuine supply chain, or illegally sold
at black and grey markets. 

Besides tracking and cloning attacks, it is also necessary to prevent replay and timing attacks, and to 
provide forward security for a secure RFID authentication protocol. Replaying the messages exchanged between a tag and a reader, and computing the time from the responses of tags, an
adversary should not be able to extract any important information about the tag. Moreover, even if an adversary can reveal the
secret of a tag in a particular session, it should not be able to compute the secrets of previous sessions; thus 
keeping forward secrecy. Again, compromise of one tag should not lead to compromise of other tags in the environment. 

There is a point of argument whether cryptography is required for low-cost tags. The emerging security concerns that we have just discussed about the RFID tags convince us to use cryptography. Of course, the resource requirement of public key cryptography is well beyond the resources available to the low-cost tags. We thus envisage the need for lightweight symmetric key cryptography for the low-cost tags. 

With an objective to design a secure and low-cost RFID authentication protocol, Gene Tsudik \cite{Tsudik} proposed a timestamp-based RFID authentication protocol using symmetric keys, named YA-TRAP$^*$. This protocol requires only lightweight hash functions and pseudo-random number generation. But this protocol is susceptible to timing attacks. 
The timestamp to be sent by the reader to the tag, can be freely selected by an adversary. Consequently, the 
adversary can mount a timing attack aimed at determining the epoch corresponding 
to the tag's last timestamp of use. Moreover, a tag can become inoperative after exceeding its pre-stored threshold timestamp value. Again, the protocol does not provide reader authentication. As we have discussed earlier, a reader needs to authenticate itself to a tag in order to prevent unauthorized tracking of the tag. 

While secure authentication is a key issue, computational and communication complexity in batch-mode authentication are the two other prime factors related to performance of an RFID system where the tags are highly resource-constrained. In batch-mode, a reader scans numerous tags, collects
the replies, and sometimes, performs identification and
authentication later in bulk. The batch-mode is appropriate
when circumstances prevent or inhibit contacting the back-end
server in real time. An inventory control system, where readers
are deployed in a remote warehouse and have no means of
contacting a back-end server in real time, is such an application. More generally, some of
the following factors might prompt the use of the batch-mode:

- The server is not available in real time, either because it is down, disconnected
or because readers do not have sufficient means of communication.

- The server is available, but is over-loaded with requests, causing response
time to be jittery, thus making each tag interrogation instance unacceptably
slow.

- The server is available and not over-loaded but is located too far away,
causing response time to be too long. (Or, the network is congested, which
cause unacceptable delays).

- A mobile/wireless reader has limited resources and, in order to conserve
battery power, simply can not afford to contact the server for each scanned
tag.

When there many tags to be authenticated at one time, it is important to do so with a communication-efficient way. Thus the need for a secure and efficient RFID authentication protocol is of paramount importance. The required security properties must be realized despite the fact that the low-cost tags are highly resource-constrained, and the computational and communication complexity should be addressed as well.  

$\textbf {Our Contribution:}$ In this paper, we introduce two mutual authentication protocols based on symmetric key where both tag and reader
authenticate each other. Our protocols achieve several security properties as they provide 
resistance against cloning, replay, tracking, timing attacks. They provide forward security, and do not allow a tag to become inoperative unless 
explicitly disabled by the legitimate reader. As for the cost, the required computation by a tag is kept at a minimum. By introducing aggregate function for the reader-to-server communication, the communication cost is reduced, which is desirable for batch-mode. In one of our protocols, the reader uses partial authentication to 
keep the suspected rogue tags out of the aggregate function, hence avoiding the failure of function verification.  Compared to YA-TRAP$^*$, our work has the following significant advantages:

\begin{itemize}
\item Provides reader authentication.

\item Thwarts timing attack.

\item Can renew threshold $T_{max}$ of time-stamp to recover from inoperative state.

\item Utilizes aggregated hash value $H$ to reduce the communication cost from reader to server.

\item Reduces the communication cost from tag to reader.

\item Comes up with formal definitions of the achieved security properties and provides security proofs based on assumptions.

\item Has been shown to be capable of being implemented under today's available RFID technology.
\end{itemize}
Our work is thus robust, secure and efficient. 

$\textbf {Paper Organization:}$ The remainder of the paper is organized as 
follows: Section
2 describes operating environment and definitions. Section 3 includes related previous work. Next, our schemes and their achievements are discussed in Section 4.
Section 5 and 6 contain the security and performance analyses, respectively. Then follows Section 7 with concluding remarks.


\section{Operating Environment}
In this section, at first, we discuss our assumptions and the operating 
environment. Then we define the privacy and security threats that should be 
resisted by an RFID system. 

\subsection{RFID System and Our Assumptions}
In an RFID environment, an RFID system consists of three
components: tags, reader(s) and server.

A tag consists of an integrated circuit with a small antenna,
and it is placed on objects that should be identified (e.g. medicine packaging, etc). Each tag will send its
 corresponding information
when interrogated by a valid reader.

Reader(s) communicate with a server and with the tags.
They are responsible for making queries to the tags.
They also have computational and storage capabilities.

A server is a trusted entity that knows and maintains all
information about tags, such as their assigned keys. A server
is assumed to be physically secure and not subject to attacks.
Multiple readers might be assigned to a single server. A server
only engages in communication with its constituent readers.
All communication between server and reader is assumed to
be over a private and authentic channel. We assume that an
adversary can be either passive or active. It can corrupt or
attempt to impersonate or incapacitate any entity, or track
RFID tags. An adversary succeeds in tracking a tag if it
has a non-negligible probability to link multiple authentication
and/or state update sessions of the same tag. Compromise of
a set of tags should not lead to the adversary's ability to track
other tags. 

Both reader and server have ample storage and computational
capabilities. We assume that an RFID tag has no clock, but has small amounts of ROM to store a key, and non-volatile
RAM to store temporary timestamps. With power supplied by the
reader, a tag can perform a modest amount of computation like bit-wise XOR, concatenation, one-way hash function, 
random number generation etc. A tag can also change the permanent state information stored in its memory. Such tags 
have been largely considered in recent literature \cite{avoine,juels,molner,MolnerWagner,squash,Tsudik}. 
The protocol transcripts are fully accessible by
the adversary.

\textbf{Notations}

Table 1 presents the notations used in this paper.

\begin{table*}[h]
\caption{Notations Used in This Paper}

\hbox to\hsize{\hfil
\begin {tabular}{|l|l|}
\hline
 $Hash$& a one-way hash function    \\
 \hline
 $H$& an aggregate hash function    \\
\hline
$k^j_i$ & a secret key of a tag $i$ at time $j$\\
\hline
$T_0$ & the initial timestamp assigned to a tag\\
\hline
$T^j_{r_i}$ & timestamp generated by the reader for tag $i$ at time $j$\\
\hline
$T^{j-1}_{t_i}$ & timestamp stored by the tag $i$ at time $j$\\
\hline
$T_{max_i}$ & the highest possible timestamp, a secret value for a tag $i$.\\
\hline
$PRNG^j_i$ & the $j$-th invocation of the PRNG of tag $i$\\
\hline
$R^j_{r_i}$ & random number generated by the reader at time $j$ for tag $i$\\
\hline
$R^j_{t_i}$ & random number generated by the tag $i$ at time $j$ \\
\hline
$R^l_{t_i}$ & random number generated by the tag $i$ at time $l$ ($l < j$)\\
\hline
$R^j_{t}$ & concatenation of random numbers generated by $n$ tags at time $j$\\
\hline
$H_{id_i}$ & a MAC value generated by tag $i$ to be authenticated by a reader \\
\hline
$AT^j_{t_i}$ & authentication token generated by tag $i$ at time $j$\\
\hline
$AT^j_{s_i}$ & authentication token generated by the server for tag $i$ at time $j$\\
\hline
$MSG$ & aggregate function verification result sent by server to reader, with \\
      &`TAG-VALID'/`TAG-AUTH-ERROR' denoting success/failure.\\
\hline
$\oplus$ & bit-wise XOR operation\\
\hline 
$||$ & concatenation of two strings\\
\hline
\end{tabular}\hfil}

\end{table*}

A keyed-hash function is used in our protocols. Such hash functions can be used to generate MAC (Message Authentication Code) values \cite{bellare}. The hash function used in our protocols are one-way and its output is pseudo-random, we also use the random
oracle. This hash function can also be used for aggregation \cite{zhu,amac}. We also assume that the random numbers used in the protocol are 64-bits in size. As for the secret key, in practice, a 64-bit key $k^j_i$ will suffice. $T_0$ can be the 
timestamp of manufacture. $T_0$ need not be tag unique;
an entire batch of tags can be initialized with the same
value. $T_{max_i}$ can be changed when a tag becomes inactive due to exceeding the
value. Even if the timestamp covers up to 1000 years
with a precision down to $nsec$, this can be covered
with 64 bits, which is much less than a full word size (e.g.,
163 bits) for a reasonable security level \cite{ecrac4}.

\textbf{Procedures.} A RFID scheme is composed of the following procedures, where
$s$ is a security parameter.

\begin{itemize}
\item{SetupServer($1^s$)} is a probabilistic algorithm which generates a key $k$, maximum allowable timestamp $T_{max}$, and timestamp $T_s$
for the server. 
\item{SetupTag($i, k_i, T_{max_i}$)} is a probabilistic algorithm which returns a tag-dependent
secret key $k_i$ and maximum allowable timestamp $T_{max_i}$. $(i,k_i, T_{max_i})$ is added in the server's database $D_S$ containing
the whole set of legitimate tags ( note that $T_t = T_s$).
\item{Auth} is an interactive protocol $\pi$ between the Server $S$ taking as inputs $k_i$, $T_{r_i}$ and $D_S$, and a tag $i$ taking as inputs $k_i$, $T_{max}$, $T_{t_i}$. At the end, the server either accepts the tag and outputs MSG=TAG-VALID or outputs MSG=TAG-AUTH-ERROR.
\end{itemize}

\textbf{Definition of the adversary.} In all experiments given below, a challenger $C$
initializes the system and the probabilistic polynomial time (PPT) adversary is given the access to some oracles.
We distinguish in the following legitimate tags from corrupted one for which
the adversary knows the secrets embedded in it. Moreover, the adversary plays
any role in the protocol by e.g. deleting or modifying some requests or
responses. We assume that an
adversary can be either passive or active. It can corrupt or
attempt to track RFID tags. More precisely, in all below experiments, $A$ has access to the following oracles.

- $O^{CreateTag}(i)$: add a tag to $D_S$ with unique identifier $i$ and key $k_i$.

- $O^{Corrupt}(i)$: returns $k_i$ and flags this tag as corrupted.

- $O^{Launch}()$: makes the server launch the first request of a new Auth protocol
instance $\pi$.

- $O^{SendReader}(m, \pi)$: sends a message $m$ to the reader for the protocol $\pi$ and
outputs the response $r$.

- $O^{SendTag}(m, i)$: sends a message $m$ to tag $i$ and outputs its $r$.

- $O^{Return}(\pi)$: outputs the result of the protocol $\pi$, that is 0 if the output of the
server during $\pi$ is 0 and 1 otherwise.

- $O^{Execute}(i)$: executes a complete Auth protocol between the server and the
tag $i$. Its output is the one of the $O^{Return}$ oracle (1 if accepted and 0 is
rejected) together with the transcript of the protocol.

\subsection{Security Definitions}
We consider the following security properties that should be achieved by a privacy-preserving mutual RFID authentication protocol. These security properties have largely been considered by RFID security research community \cite{cloning1,burmester,AutoID,dot,juels,cloning2,burmester1,replay,song,sm:bp}. We assume active adversarial threat against forward security and backward security features. Security properties like tracking, timing, replay, desynchronization are assumed to be feasible for a passive adversaries. Cloning attack can be launched by either an active or passive adversary.

\textbf{Forward Security:} If an adversary compromises an entity, then it might
be able to derive previous keys to track old transactions involving that 
entity, thus violate forward security. Bellare et.al. \cite{bellare:forward} demonstrated forward security for symmetric key cryptosystems. 
They proposed a forward secure pseudo-random
bit generator, and showed the construction of forward-secure message authentication schemes where a server is also subject to compromise. However, our definition follows \cite{song} for
 forward security, where it is assumed that the server is not subject to compromise. The goal of \cite{song} and \cite{bellare:forward} has been to achieve forward security such 
 that all past secret keys are secure when the current secret key is exposed. The definition below follows the one presented in \cite{fordsec}.
 
\textit{Forward Security experiment}

1. $A$ executes oracle queries except $O^{Corrupt}, O^{CreateTag}$ for all $n - 1$ tags, except for the tag $i_c$ used in challenge phase.

2. $A$ selects a challenge tag $i_c$ from the set of $n$ tags, and executes oracle queries except $O^{CreateTag}$ for $i_c$'s $j$-th instance.

3. $A$ calls the oracle $O^{Return}$ for $j-1$-th instance, and challenger $C$ tosses a fair coin $b \in_R \{0, 1\}$ s.t. if $b = 1$, $A$ is given the messages corresponding to $i_c$'s $(j-1)$-th instance, else is given random values.

4. $A$ executes the oracles for $n - 1$ tags, except $i_c$, like in the learning phase (step 1).

5. $A$ outputs a guess bit $b'$, and it wins if $b = b'$

\textit{Definition (Forward Security)}: We say that an RFID scheme has the forward security property if the probability that $b = b'$ differs from $1/2$ by a fraction that is at most negligible.

\textbf{Backward Security:} This can be defined similarly as forward security where knowledge of a tag's internal state at time $j$
can help to identify tag interactions that occur at a time $j'> j$ \cite{lk,sm:bp}. Since the adversary is able to trace the target tag at least during the authentication immediately
following compromise of the tag secret, perfect backward security makes no sense. Therefore, a minimum restriction should be imposed to achieve backward security, such that the adversary misses the necessary protocol transcripts to update the compromised key. Although this assumption for backward security is true for certain classes of privacy-preserving RFID protocols (i.e., for shared key environment)\cite{lk,sm:bp}, it is clearly not true for some other cases. For instance, Vaudenay shows an RFID protocol based on public-key cryptography that is resistant to this attack \cite{vaud:asia}. However, our notion of backward security is true for privacy-preserving RFID protocols based on shared secrets that are updated on each interaction between tag and reader. Backward security is thus harder to achieve than forward security in general, particularly under the very constrained environment of RFID tags. However, backward security is never less important than forward security in RFID systems. In
the case of target tracing, it suffices to somehow
steal the tag secret of a target 
and collect interaction messages to trace the future behaviors of the particular target. Without backward security, this kind of target tracing is trivial. In the case of supply chain management systems, even a catastrophic scenario may take place without backward security: if tag secrets are leaked at some point of tag deployment or during their time in the environment, then all such tags can be traced afterwards. We give the definition formally below which follows the one presented in \cite{fordsec}.

\textit{Backward Security experiment}

1. $A$ executes oracle queries except $O^{Corrupt}, O^{CreateTag}$ for all $n - 1$ tags, except for the tag $i_c$ used in challenge phase.

2. $A$ selects a challenge tag $i_c$ from the set of $n$ tags, and executes oracle queries except $O^{CreateTag}$ for $i_c$'s $j$-th instance. [$A$ is not given the transcript necessary to update $i_c$'s key.]

3. $A$ calls the oracle $O^{Return}$ for $j+1$-th instance, and challenger $C$ tosses a fair coin $b \in_R \{0, 1\}$ s.t. if $b = 1$, $A$ is given the messages corresponding to $i_c$'s $(j+1)$-th instance, else is given random values.

4. $A$ executes the oracles for $n - 1$ tags, except $i_c$, like in the learning phase (step 1).

5. $A$ outputs a guess bit $b'$, and it wins if $b = b'$

\textit{Definition (Backward Security):} We say that an RFID scheme has the backward security property if the probability that $b = b'$ differs from $1/2$ by a fraction that is at most negligible.

\textbf{Tracking Attack}: If an entity's responses are linkable to each other, or
distinguishable from those of other entities, then the entity's location could 
be tracked by an adversary \cite{song, sm:bp}. For example, if
the response of a tag to a reader query is a static ID
code, then the movements of the tag can be monitored,
and the social interactions of an individual carrying a
tag may be available to third parties without him or
her knowing. If messages from tags are anonymous,
then the tag tracking problem can be avoided.

More formally, the goal
of the adversary $A$ is to recognize one tag among two and/or to distinguish two different responses of same tag.

\textit{Tracking Experiment 1}:

1. At any time of the game, $A$ chooses two tags $i_0$ and $i_1$ in the set of
legitimate tags and sends $(i_0, i_1)$ to $C$.

2. $C$ randomly chooses a bit $b$. The tag $i_b$ is called the challenge tag. $i_0$
and $i_1$ are withdrawn from $D_S$ and thus cannot be manipulated by the
adversary using oracles. $i_b$ is added to $D_S$, as an exact copy of the tag
$i_0$ or $i_1$.

3. Again, $A$ interacts with the whole system through all oracles. Note that $A$
can interact with the challenge tag without any restriction.

4. $A$ finally outputs a bit $b'$ and wins if $b=b'$.

\textit{Tracking Experiment 2}:

1. $A$ executes the oracles for all $n-1$ tags except tag $i_c$.

2. $A$ chooses a tag $i_c$ in the set of
legitimate tags, runs the oracles for any time $j$.

3. For $A$'s query to $o^{Return}$, $C$ tosses a fair coin $b \in_R \{0,1\}$ s.t. if $b=1$, $A$ is given the messages corresponding to $i_c$'s $j$-th instance, otherwise a random value is given.

4. $A$ finally outputs a bit $b'$ and wins if $b=b'$.

\textit{Definition (Tracking resistance:)} We say that an RFID scheme has the tracking resistance property if the probability in Tracking Experiments 1 and 2 that $b = b'$ differs from $1/2$ by a fraction that is at most negligible.

\textbf{Cloning Attack:} A cloning attack is an integrity attack in which an attacker succeeds in
capturing a tag's identifying information. In a tag cloning attack, an attacker may
install a replacement tag that emits an identifier similar to
the original one \cite{dot}. In a tag ownership transfer scenario, the old owner, acting as an active adversary, may want to clone the tag whose ownership has been changed. Since tag's old secret is known to the old owner, in such a case, the old owner may want to implant the old secret to a fake tag to make a clone of the genuine tag. While the above scenario considers active adversaries, cloning may also be possible under a pssive attack where the adversary tries to extract the tag's secret information from the protocol transcripts \cite{burmester}. A direct consequence of cloning is the possibility of counterfeiting, where a genuine RFID-tagged article may be reproduced as a cheap counterfeit and tagged with a clone of an authentic RFID tag. This would fool the system into believing
the product is still on the shelf, or alternatively, an expensive
item could be purchased for the price of a cheap one. These
types of attacks can have serious consequences. Examples of
such attacks have been demonstrated in various works \cite{cloning1,cloning2}.
The ability to create clones of tags exposes corporations to new vulnerabilities if RFIDs are
used to automate verification steps to streamline security procedures \cite{burmester}. 

More formally, a fake tag cannot be accepted by the
system. It corresponds to the strong soundness in \cite{damgard} where the adversary can
corrupt tags. The following experiment considers active adversarial model. Experiment for passive model can be designed in a similar fashion by not allowing the adverasry to access $O^{Corrupt}$ oracle.

\textit{Cloning Experiment:}

1. At any time of the game, $A$ makes successful calls to the  $O^{Corrupt}$, $O^{Launch}$, $O^{SendReader}$, $O^{SendTag}$ oracles, and a final call to the $O^{Return}$ oracle.

2. The experiment's output is 1 if $A$ is accepted during the Auth protocol and
the outputted tag is not corrupted, and 0 otherwise.

\textit{Definition(Cloning):} An RFID system resists cloning attack if the probability that
the bit $b$ returned by the $O^{Return}$ oracle at the end of the Cloning Experiment
is equal to 1 is negligible.

\textbf{Replay Attack:} An adversary could intercept messages exchanged between a
reader and a tag, and replay them. In such an attack, an attacker reuses
communications from previous sessions to perform a
successful authentication between a tag and a server \cite{replay,sm:bp}.

\textit{Replay Attack Experiment:}

1. At any time $j$ of the game, $A$ plays the role of a tag in the protocol by
using successful calls to the $O^{SendReader}$, $O^{SendTag}$ oracles, and a final call to the $O^{Return}$
oracle.

2. At any time $j'$ ($j \neq j'$), $A$ queries $O^{SendReader}$,$O^{Execute}$, $O^{Return}$ oracles by using the outputs of $O^{SendReader}$, $O^{SendTag}$ oracles of time $j$. The experiment's output is 1 if $A$ is accepted during the protocol and the outputted tag is not corrupted, and 0 otherwise.

\textit{Definition(Replay):} An RFID system resists replay attack if the probability that
the bit $b$ returned by the $O^{Return}$ oracle at the end of the Replay Experiment
is equal to 1 is negligible.

\textbf{Timing Attack:} The attacker attempts to compromise a system by analyzing the 
time taken to execute cryptographic algorithms. The attack exploits the fact 
that every operation in a computer takes time to execute. Timing attack extracts
information based on variations in the rate of computation
of a target device \cite{juels}, and can launch side-channel attacks \cite{sidechannel} to get the internal data in a tag memory. Before formally defining the timing attack resistance property, we introduce an oracle as follows:

$O^{Tr}$: At any authentication session, it takes `S' (success) or `F' (failure) as the input and outputs the tag's response (transcripts) generated upon the successful or failed authentication. The oracle also returns the response time of the tag.

\textit{Timing Attack experiment} 

1. $A$ executes queries on the $O^{Tr}$ oracle on both S and F.

2. Challenger $C$ tosses a fair coin $b \in_R \{0, 1\}$ s.t. if $b = 1$, $A$ is given the transcripts and time of S, else is given the transcripts and time of F.

3. $A$ executes queries on the $O^{Tr}$ oracle as in step 1.

4. $A$ outputs a guess bit $b'$, and it wins if $b = b'$ 

\textit{Definition (Timing Attack:)} We say that an RFID scheme has the resistance against timing attack if the probability that $b = b'$ differs from $1/2$ by a fraction that is at most negligible.

\textbf{Tag Desynchronization:} Tags can get desynchronized from the server if the protocol message never reaches the tag,
say because of transmission problems or even worse because of an adversary blocking some message \cite{AutoID}. When an adversary targets major distribution centers or customs, e.g. at harbors or airports, desynchronization may severely slow down inspection processes and thus interfere with the free flow of goods.

To have a better characterization, we will distinguish on one
side the maximum number of desynchronization $Desync_S$ an adversary can create
when only focusing on the server, and on the other side this maximum number
$Desync_T$ when $A$ only focuses on the tag. The desynchronization value consequently
corresponds to the couple ($Desync_S,Desync_T$). In fact, in an experiment, let $A$ chooses a tag $i$. We
denote by $TK_i = k^j_i$ (resp. $SK_i = k^{j'}_i$) the tag's version (resp. server's
version) of $k_i$ at the end of this step. The maximum number of desynchronization obtained by the adversary $A$, when $A$ only focuses on the tag (resp. the server), is $Desync_{T,A} = j'-j$ (resp. $Desync_{S,A} = j-j'$). Our definition follows the one presented in \cite{desync:def}.

Let us denote by $Resync_S$ (resp. $Resync_T$ ) the maximum number of desynchronizations the
scheme can tolerate to accept the tag $T$ during the Auth procedure, if only the tag
(resp. the reader) has been updated, i.e. after the Auth procedure $SK_i = k^j_i$(resp. $TK_i = k^{j'}_i$). We now give methods to compute respectively $Resync_S$ and $Resync_T$. Let us consider a tag $T$ , the server $S$ (synchronized with $T$ ) and a counter $Ct$ which will be incremented in each round of the two experiments. We first introduce two procedures:
$UpdateTag(i)$, which forces $TK_i$ to be updated, and $UpdateServer(i)$ which
forces $SK_i$ to be updated. The computation of $Resync_S$ (resp. $Resync_T$ ) works as follows. During round $Ct$, we
produce $Ct$ desynchronizations of the tag (resp. the server) using $UpdateTag(i)$ (resp. $UpdateServer(i)$) and then launch the Auth procedure between $S$ and $T$ .
If the server accepts the tag, $TK_i$ and $SK_i$ are resynchronized (i.e. $TK_i = SK_i$), $Ct$ is incremented and a new round is started, the tag (resp. the server) will be updated once more. Else we stop the algorithm
and output the value $Ct-1$ which is exactly $Resync_S$ (resp. $Resync_T$).

\textit{Definition (Synchronization:)} For a given RFID authentication scheme, the desynchronization
value of a scheme is the couple ($Desync_S,Desync_T$) with $Desync_S = Sup_A(Desync_{S,A})$ and $Desync_T = Sup_A(Desync_{T,A})$. The scheme is said ($Desync_S,Desync_T$)-desynchronizable. Also, the resynchronization value of the scheme is the couple ($Resync_S,Resync_T$) and the scheme is said ($Resync_S,Resync_T$)-resynchronizable. If $Desync_S \leq Resync_S$ and
$Desync_T \leq Resync_T$ , the scheme is said synchronizable. Else, the scheme is said desynchronizable.

\section{Previous Work}

To achieve forward security, 
tracking resistance, and timing attack resistance with low computational cost, \cite{rep, molner, Tsudik} use symmetric keys between a tag and a reader/server. MSW (Molner, Soppera, Wagner) protocol \cite{molner} 
uses hierarchical tree based keying to provide efficient tag authentication. However, the amount of
computation required per tag is not constant, but logarithmic
with the number of tags in the hash-tree. Also, MSW protocol has a security flaw whereby an adversary who 
compromises one tag, is able to track/identify other tags that belong to the 
same families
(tree branches) as the compromised tag \cite{avoine}. Finally, MSW scheme does not provide forward security \cite{burmester1}. REP \cite{rep} has been proposed to satisfy security 
properties like forward security, and  resistance against cloning, timing, tracking with low computational cost. But the main drawback of REP is that a tag needs to be attached with an 
additional or proxy device to be able to communicate and to do the necessary computations for authentications. 
Again, the purpose of timestamp-based YA-TRAP* \cite{Tsudik} has been to achieve resistance against cloning and tracking attacks, and forward security with low cost. 
But it is susceptible to $timing$ $attacks$. 
The timestamp $T_r$, to be sent by the reader, can be freely selected by an adversary. Consequently, the 
adversary can mount a timing attack aimed at determining the epoch corresponding 
to the tag's last timestamp of use ($T_t$) \cite{Tsudik}. Moreover, in YA-TRAP$^*$, a tag can become 
inoperative after exceeding the pre-stored threshold timestamp value, and it does not provide reader authentication.

On the other hand, to achieve the security features like forward security, and resistance against cloning and tracking attacks, public key cryptography has been used by a 
series of EC-RAC protocols \cite{ecrac3,ecrac,ecrac2}. These are provably secure RFID authentication protocols based on 
ECDLP. These were shown to be vulnerable to tracking and replay attacks by \cite{Bringer,fan,ton}. However, a new authentication scheme \cite{ecrac4} has been proposed addressing the vulnerabilities of the previous versions. Nevertheless, this protocol requires
 aroung 14,500 gate counts (NAND gate 
equivalent) and an EC processor, which are well beyond the capability of today's low-cost passive RFID tags.

Communication cost is another important issue in an environment where many tags are read at the same time (for example, 
supply chain, inventory management, etc.). Zhu et.al. \cite{zhu} showed the security of aggregate functions for RFID 
tags which focuses on reducing communication cost. 
The use of aggregate functions reduces communication complexity, which is a prime factor related to lowering energy 
consumption of an RFID system. But 
they do not show the use of the aggregate function in a full authentication 
protocol. Moreover, it is not clear from their work how the server can find rogue/fake tags.

\section{Our Low-Cost and Secure Scheme}\label{sec:item}

Considering the above discussions, our purpose is to improve YA-TRAP$^*$ by providing the properties like reader authentication and resistance against timing attack, and allowing a tag to renew its pre-stored timestamp to recover from inoperative state. Our purpose is to reduce the communication cost, too. We propose two schemes to serve our purpose. The first scheme reduces communication costs from tag to reader, and a enables a server to authenticate many tags at once without partial authentication (partial authentication has been described in the following subsection). 
In our second scheme described in Subsection 4.3, we extend the first scheme into a more secure one by providing partial authentication of the tags in batch-mode. In the extended version, 
a tag's computation increases by one hash function, and the tag-to-reader 
communication cost increases by $b$ bits, where $b$ is the bit-length of the protocol messages. 

In the initial phase, both a server and a reader authenticate each other. We do not consider the server-to-reader
communication cost of the initial phase/set up phase when they authenticate each other, and when the server sends the required tag-related information to the reader. Also, the initial secrets (e.g. shared secret key $k_i^0$, maximum timestamp $T_{max_i}$) can be given to a tag in the setup phase. The setup phase can take place during the manufacturing time. All these communication can be done off-line. We are mainly concerned about the on-line
communication costs.

\subsection{Protocol Building Blocks}

$\textbf {1) One-time pad}$: The one-time pad is a simple, classical form of encryption (See, \cite{menezes} for 
discussion). We briefly restate the underlying idea. If two parties share a secret one-time pad $p$, for example a
 random bit string, then one party can transmit a message $m$ secretly to the other via the ciphertext $p \oplus m$,
  where $\oplus$ denotes the XOR operation. It is well known that this form of encryption provides unconditional 
  security. Suppose, for instance, that pads from two different verifier-tag sessions are XORed with a given tag
value in order to update the tag. Then, even if an adversary intercepts the pad used in one session, he/she will learn 
no information about the updated value. Application of a one-time pad requires only the lightweight computational 
process of XORing. The standard cryptographic primitives require 
more computational power than one-time pads. This is why we use one-time pads.\\
$\textbf {2) Aggregate function}$: An aggregate function compresses the size of all hash functions $Hash$, so that the 
communication complexity between the reader and the server can be reduced accordingly. This leaves an interesting 
research problem - is it possible to aggregate tags' attestations so that the size of the resulting aggregate 
attestation is approximate to that of the original case (i.e., non-aggregate model)? 
That is, given $H_{id_i}= Hash(R^j_{t_i} \lVert R^j_{r_i}, k_i^j)$, where $Hash$ is a one-way hash function, 
and $R^j_{t_i}$ and $R^j_{r_i}$ are random challenges of tag and reader respectively at time $j$, we ask whether 
there exists an efficient polynomial time algorithm such that on input $H_{id_i}= Hash(R^j_{t_i} \lVert R^j_{r_i}, k_i^j)$,
 it outputs an aggregate of hash functions $H=\bigoplus_{i=1}^{n}  H_{id_i}$ whose size is approximately 
 the same as an individual $H_{id_i}$. Moreover, the validity of individual attestations can be checked efficiently 
 given the aggregate hash $H=\bigoplus_{i=1}^{n}  H_{id_i} $.

We derive an aggregate function from \cite{zhu} as a tuple of probabilistic polynomial time algorithms 
(Hash, Aggregate, Verify) such that:

The authentication algorithm Hash takes as input random numbers $R^j_{t_i}$ and $R^j_{r_i}$, and key $k_i^j$. 
The algorithm outputs an attestation $H_{id_i}$;

The aggregate function Aggregate takes as input ($H_{id_1},H_{id_2}, ... ,H_{id_n}$) and outputs a new attestation $H$;

The verification algorithm Verify takes as input ($R^j_{t_1} , R^j_{r_1}$, $k^j_1$),
($R^j_{t_2} , R^j_{r_2}, k^j_2$),..., ($R^j_{t_n} , R^j_{r_n}, k^j_n$) and an attestation $H$, outputs a MSG, 
with TAG-VALID denoting acceptance or TAG-AUTH-ERROR denoting rejection.\\
$\textbf {3) Security notions based on aggregate function}$: Our protocols use authentication in batch-mode either with or without partial authentication. Authentication in batch-mode requires that more than one tags are authenticated at once. The use of aggregate function does not rule out the possibility that some rogue/illegal tags' information is included in the function. The aggregate function only enables a server to find out an anomaly in the 
resultant XOR operations of the hash values. This means, if the computed 
aggregate hash value does not match with the received aggregate hash value, the server cannot 
identify the specific tag for which the result is an oddity given that a protocol does not come up with a partial authentication. Partial authentication helps an aggregate function to be correctly 
verified by the server, hence authenticating the corresponding tags. The notion of partial authentication is such that, the server computes authentication tokens for each tag, and sends the tokens to the reader before an authentication session takes place. During an authentication session, a tag computes its authentication token, and sends the token to the reader. The reader then matches the newly received token with the stored one. If they do not match, the reader excludes the tag from the input of aggregate function. Thus, partial authentication ensures that suspected rogue tags are out of the aggregate function computation, hence providing a more secure batch-mode authentication than that without partial authentication.

\subsection{Our Basic Scheme}

At any session $j$, a tag $i$ has ($T^{j-1}_{r_i}, T_{max_i}, k_i^j$) in its memory. Similarly, for each tag $i$, the server has ($T^{j-1}_{r_i}, T_{max_i}, k_i^j$) stored in its memory.

Algorithm 1 below is our secure and low-cost authentication
protocol:
  \begin{algo}[Low-Cost and Secure Scheme]~\par
 $[1] Tag \leftarrow$ Reader: $T^j_{r_i}, R^j_{r_i}, Hash(T^{j-1}_{r_i} \parallel T^j_{r_i}, T_{max_i})$

 $[2]$ $Tag_i$:
 
\indent $\hspace{0.3cm}$ $[2.1]$While $Hash (T^{j-1}_{t_i} \parallel T^j_{r_i}, T_{max_i}) \neq Hash 
(T^{j-1}_{r_i} \parallel T^j_{r_i}, T_{max_i})$,

\indent $\hspace{0.4cm}$ $R^j_{t_i}= {PRNG_i}^1 $, $H_{id_i}= {PRNG_i}^2 $, $k^{j+1}_i= {PRNG_i}^3 $

\indent $\hspace{0.3cm}$ $[2.2]$While $T^j_{r_i} > T_{max_i}$, then $T_{{max_i}_{new}}= T^j_{r_i} \oplus T_{max_i}$;

\indent $\hspace{0.4cm}$ if $T_{{max_i}_{new}} > T_{max_i}$ , then 
 set $T_{max_i} = T_{{max_i}_{new}}$, 

\indent $\hspace{0.5cm}$else $R^j_{t_i}= {PRNG_i}^1 $, $H_{id_i}= {PRNG_i}^2 $, $k^{j+1}_i= {PRNG_i}^3 $

\indent $\hspace{0.3cm}$ $[2.3]$ $\delta = T^j_{r_i} - T^j_{t_i}$

\indent $\hspace{0.3cm}$ $[2.4]$ if $(\delta \leq 0) $ then

\indent $\hspace{0.4cm}$ $R^j_{t_i}= {PRNG_i}^1 $, $H_{id_i}= {PRNG_i}^2 $, $k^{j+1}_i= {PRNG_i}^3 $

\indent $\hspace{0.4cm}$ else $T^j_{t_i} = T^j_{r_i}$, $R^j_{t_i} = PRNG_{i}$, $H_{id_{i}} = Hash(R^j_{t_i} \lVert R^j_{r_i}, k^j_i)$ 

\indent $\hspace{0.3cm}$ $[2.5]$ $k^{j+1}_i = Hash(k^j_i, R^j_{r_i})$

 $[3]$ Tag $\rightarrow$ Reader: $H_{id_i},R^j_{t_i} $ 
 
 $[4]$ Reader:
 
\indent $\hspace{0.3cm}$ $[4.1]$ If $R^j_{t_i}$ matches any of the previously generated $R_{t_i}$, then REJECT

 \indent $\hspace{0.4cm}$ else mark each $ H_{id_i}, R^j_{t_i}$ 
 
\indent $\hspace{0.3cm}$ $[4.2]$ $H=\bigoplus_{i=1}^{n}  H_{id_i} $, $R^j_t= R^j_{t_1} \parallel R^j_{t_2} \parallel R^j_{t_3} 
\parallel ..... \parallel R^j_{ t_n} $ 

 $[5]$ Reader $\rightarrow$ Server: $H, R^j_t$
 
 $[6]$ Server:
 
\indent $\hspace{0.3cm}$ $[6.1]$  lookup accepted $T_{r_i}$ according to marked $R_{t_i}$;

\indent $\hspace{0.5cm}$if $ \bigoplus_{i=1}^{n} Hash(R^j_{t_i} \lVert R^j_{r_i}, 
k^j_i) \neq H $, then MSG=TAG-AUTH-ERROR

\indent $\hspace{0.4cm}$ else MSG=TAG-VALID

\indent $\hspace{0.3cm}$ $[6.2]$ update each $k^j_i$

 $[7]$ Server $\rightarrow$ Reader: MSG

\end{algo}

In the search procedure, for each tag $i$, the server computes the value $Hash(R^j_{t_i}\lVert R^j_{r_i}, k^j_i)$ and compares it with $H_{id_i}$ (in case of more than one tag, the values are aggregated and compared with the aggregate value $H$). In case of a match, the procedure
outputs MSG=TAG-VALID, and updates the key $k^j_i$. Else, for each $i$, the server computes the ephemeral key $Ek_i^{j+1} = Hash(k_i^j, R^j_{r_i})$, computes $Hash(R^j_{t_i}\lVert R^j_{r_i}, Ek^{j+1}_i)$ and compares it with $H_{id_i}$. In case of a match, the server outputs MSG=TAG-VALID and updates and updates the key $k^j_i$ (so the key updated in the tag = key updated in the server). Otherwise, the tag is rejected.

Note that, the tag rejects any illegal reader by 
generating pseudo-random numbers(PRNG). PRNGs are used to generate `reject' messages in order to consume the same 
computation time as computing a hash function. The use of PRNGs to hide a tag's identity was first introduced in 
\cite{Weis}.

The server keeps the table of valid tags and the last issued timestamp 
$T_{r_i}$ for each of the tags, so, it can easily find a valid tag which 
becomes inoperative. The new timestamp threshold value $T_{{max_i}_{new}}$ is 
stored in memory, erasing the existing $T_{max_i}$ only after the reader 
authenticates itself to the tag. However, $T_{{max_i}_{new}}$ is generated only when previous $T_{max_i}$ has been exceeded (unexpected output from a valid tag can be an 
indicator). The reader sends the 
value $T^j_{r_i}$ by XOR-ing the $T_{max_i}$ and $T_{{max_i}_{new}}$. Generating 
$T_{{max_i}_{new}}$ works as one-time padding, as it is freshly computed every 
time a tag is required to update its $T_{max_i}$. Before XOR-ing, the reader must make sure 
that the value of $T_{{max_i}_{new}}$ is strictly greater than $T_{max_i}$.


\subsection{Extending the Protocol}

The aggregate function only enables a server to find out an anomaly in the 
resultant XOR operations of the hash values. This means, if the computed 
aggregate hash value does not match the received $H$, the server cannot 
identify the specific tag for which the result is an oddity.

To reduce such incidents, we use a one-way hash for the partial authentication 
of a tag. One-wayness means that, having seen the hash value, it is not possible 
to extract the contents of the hash. For this purpose, we need to add the 
following step to compute an authentication token(AT) as Step [2.5] before 
renewing the key $k^j_i$:

$[2.5] AT^j_{t_i}= Hash(T_{max_i}, k^j_i)$\\
A tag $i$ has its secret value $T_{max_i}$ and key $k^j_i$. Generating $k^j_i$ 
for every read operation also ensures forward security. So, to partially 
authenticate itself to server, a tag sends the computed $AT^j_{t_i}$ value to 
the reader. So, Step [3] is rewritten as:

$[3]': Tag \rightarrow Reader: H_{id_i},R^j_{t_i},AT^j_{t_i}$\\
Upon receiving $AT^j_{t_i}$ from a tag $i$, the reader finds a match
with the desired $AT^j_{s_i}$ [the value $AT$ computed by the server for a tag 
$i$ at time $j$] for tag $i$, which the reader had received during ($j-1$)-th session from the server for that particular tag $i$ for the $j$-th read interaction. 
This requires us to modify Step [4.1] as [4.1]', shown below:

$[4.1]'$ If $R^j_{t_i}$ matches any one of the previously generated $R_{t_i}$, 
then REJECT

else if $AT^j_{t_i}\neq AT^j_{s_i}$, then exclude tag $i$

else mark each $AT^j_{t_i},H_{id_i},R^j_{t_i}$\\
The reader has a table consisting of the expected identification token 
$AT^j_{s_i}$ values corresponding to each tag (each tag's $T^j_{r_i}$). So, the 
reader checks the hash value of a tag, and if a tag is found to be legitimate, its 
authentication message is included into the reader's aggregate function.

The tag sends a one-way hash value to the reader to 
authenticate itself. As the reader is not capable of too many computations, such as 
computing hash values of a huge number of tags, it forwards the hash values for 
authentication to the server.

The server also needs to do some computations after it matches the aggregated 
value received from the reader. After updating each key $k^j_
i$ of the corresponding tags to $k^{j+1}_i$, the server computes their 
respective authentication token values ($AT^{j+1}_{s_i}$) for the next 
($j+1$)-th read operation. This step is added to the protocol as Step [6.3]:

$[6.3]$compute $AT^{j+1}_{t_i} = Hash(T_{max_i}, k^{j+1}_i)$

Furthermore, the server-to-reader communication must include the newly 
generated $AT^{j+1}_{s_i}$ values. So, Step [7] can be modified as:

$[7]'$ Server $\rightarrow$ Reader: MSG, all $AT^{j+1}_{s_i}$


One significant point here to notice is that the extended version of the protocol 
increases communication cost for both the tag-to-reader and server-to-reader 
interactions. However, increasing this communication cost strengthens security. We consider this trade-off between 
cost and security to be a feature 
of our protocols. The partial authentication by the reader helps filter out 
malicious tags. This eliminates inclusion of any rogue tag in the aggregate 
function, which leads to no anomaly when the server verifies the aggregate 
value.

We emphasize that, for a batch-mode environment validating a number of 
tags in a very short of time, the initial version is suitable - where the 
communication cost is the least. In such a batch-mode environment, tags are 
authenticated in bulk, hence it is sufficient to verify the authentication of the 
whole batch together in the least possible time. Even though the extended protocol 
is not very effective for batch-mode environment, where validating a number of 
tags in a short time is required, it can be used in settings where 
the server is not readily available (Police checking drivers' licenses with 
mobile readers is such a condition, where the bulk of the licenses are finally 
authenticated later by the server; or, an inventory control system where readers 
are deployed in a remote warehouse and have no means of contacting a back-end 
server, for another example). In such situations, the reader can partially authenticate 
the tags. Moreover, the extended version is also suitable for a small number of 
tags, or even for individual
tag authentications.

\section{Security Analysis}

In this section, we give the proof sketches of the security properties.

\begin{theorem}{Forward Security:}
The scheme is forward secure under the security of the hash function.
\end{theorem}
\textit{Proof}. For each valid read operation, a tag uses the current key $k^j_i$ for creation and 
verification of MAC. At the end of each valid read operation, 
$k^j_i$ is updated by a one-way hash function $Hash$, and previous $k^j_i$ is deleted from the tag's memory. An attacker breaking into the tag's memory gets the current key. But given the current key $k^j_i$ it is still not possible to derive any of the 
previous keys due to the one-wayness of the hash function. Again, since the used hash functions are one-way, the protocol transcripts used in session $j$ are useless for $A$ even if it knows the secrets of session $j+1$. In all the above cases, the adversary has to guess the previous key correctly with a success probability of $2^{-l}$ where $l$ is bit length and is polynomial in security parameter $s$, which is negligible. So, our protocols have the property of forward security. $\square$

\begin{theorem}{Backward Security:}
The scheme is backward secure under the assumption and security of the hash function.
\end{theorem}
\textit{Proof}. Our schemes provide backward security if an adversary misses the first pass of the protocol just once
in a single successful authentication session after compromising
a tag's secret. That is, if the adversary cannot prevent a tag from receiving the transcript (i.e., $R^j_{r_i}$) that is
needed to refresh $k_i^j$, then it can compute the new
key $k^{j+1}_i = Hash (k^j_i,R^j_{r_i})$ only with a negligible success probability of $2^{-l}$ where $l$ is bit length and is polynomial in security parameter $s$. Thus, for the ($j+1$)-th session, the adversary can successfully generate $Hash(R^{j+1}_{t_i}\lVert R^{j+1}_{r_i}, k^{j+1}_i)$ with negligible probability even if it has $R^{j+1}_{t_i}, R^{j+1}_{r_i}$ from the protocol session. So, our protocols have the property of backward security. $\square$

\begin{theorem}{Tag Tracking:}
The scheme is tracking resistant under the security of one-time pad, the pseudorandomness of the hash function.
\end{theorem}
\textit{Proof}. The scheme clearly provides the anonymity of the tag under the assumption that the hash function is one-way. Moreover, the scheme is unlinkable since $A$ has no control on the value $R_{t_i}^j$ . Again, if the used hash functions are pseudorandom, the protocol transcripts are useless for $A$ since the success probability of guessing the correct secret is $2^{-64}$, which is negligible. Moreover, from the transcript $AT^j_{t_i}$ where $AT^j_{t_i} = Hash(T_{max_i}, k^j_i)$, the adversary can successfully compute $k^j_i$ and $T_{max_i}$ with the probability $2^{-l}+2^{-l}$ where $l$ is bit length and is polynomial in security parameter $s$, which is negligible. It is also not possible for an adversary to track a tag, due to the use of 
one-time padding. As the adversary cannot distinguish a normal response from a 
PRNG, he/she cannot track a tag even if the tag becomes incapacitated by exceeding the 
$T_{max_i}$ value. So when a valid reader forwards a $T^j_{r_i} > T_{max_i}$ to 
an incapacitated tag $i$, the adversary cannot find $T_{max_{i_{new}}}$ 
from the value he/she has seen during the session. As the $T_{max_{i_{new}}}$ is 
freshly generated, and the adversary cannot know for which particular value of 
$T^j_{r_i}$ the stored $T_{max_i}$ is going to be refreshed, the unconditional security of one-time padding 
provides secrecy for $T_{max_{i_{new}}}$. $\square$

\begin{theorem}{Tag Cloning:} 
The scheme is cloning resistant under the security of the hash function.
\end{theorem}
\textit{Proof}. To win the Cloning Experiment, an adversary has to first guess which random value $R^j_{r_i}$ will be sent by the server (in the case of ownership transfer scenario described in the definition). Thus, if the adversary knows the key $k^j_i$, it can guess the correct $R^j_{r_i}$ for updating the key with a success probability of $2^{-l}$ where $l$ is bit length and is polynomial in security parameter $s$, which is negligible. Another
possibility for the adversary (apassive attack) is to produce a valid message $Hash(R^j_{t_i} \lVert R^j_{r_i}, k^j_i)$ without knowing a valid (and uncorrupted) value $k^j_i$: this is not possible under the one-wayness of the hash function. So the scheme is cloning resistant. $\square$

\begin{theorem}{Timing Attack:} 
The scheme is timing attack resistant under the assumption that execution of PRNG and hash function takes same amount of time.
\end{theorem}
\textit{Proof}. The adversary can win the game if it can distinguish a tag's real response upon a seccessful session (S) from a random response upon a failed session (F). If $Hash (T^{j-1}_{t_i} \parallel T^j_{r_i}, T_{max_i}) \neq Hash 
(T^{j-1}_{r_i} \parallel T^j_{r_i}, T_{max_i})$ (server authentication fails) or $T_{{max_i}_{new}} < T_{max_i}$ (adversary feeds with arbitrary $T_{max_i}$) or $(\delta \leq 0)$ (current timestamp predates the previous one), then the tag performs ($R^j_{t_i}= {PRNG_i}^1$, $H_{id_i}= {PRNG_i}^2$, $k^{j+1}_i= {PRNG_i}^3$) instead of computing ($R^j_{t_i} = PRNG_{i}$, $H_{id_{i}} = Hash(R^j_{t_i} \lVert R^j_{r_i}, k^j_i)$, $k^{j+1}_i = Hash(k^j_i, R^j_{r_i})$) upon being successful (S). The set of operations are indistinguishable since the executions of PRNG and $Hash$ are assumed to take the same amount of time. The adversary thus has negligible probability to distinguish a successful session from a failed session. So the scheme is timing attack resistant. $\square$

\begin{theorem}{Replay Attack:} 
The scheme is replay attack resistant under the security of the hash function as a secure MAC for authentication.
\end{theorem}
\textit{Proof}. In our protocols, the reader 
matches the random numbers $R^j_{t_i}$ it receives from the tags to make sure that no two 
random numbers from two different sessions with the same tag are the same, i.e., $R^j_{t_i} \neq R^{j'}_{t_i}$ ($j \neq j'$). Moreover, to win the Replay Attack experiment, the adversary has to create a valid transcript $Hash(R^j_{t_i} \lVert R^j_{r_i}, k^j_i)$ without knowing a valid (and uncorrupted) value $k^j_i$: this is not possible under the security of the hash function assuming that the hash function is one-way and is used as a secure MAC for authentication. Moreover, the transcripts are
functions of freshly generated random numbers $R^j_{t_i}$ and $R^j_{r_i}$, and
thus these messages cannot be reused in other sessions. Also, since the timestamp must be greater than
the last-heard time of any tag, i.e., $T^j_{r_i} > T^{j-1}_{r_i}$, an adversary cannot reuse these values with the tag involved
in the communication he has eavesdropped. So, the scheme is replay attack resistant. $\square$

\begin{theorem}{Desychronization Attack:} 
The scheme is ($1,0$)-desynchronizable, ($1,0$)-resynchronizable. The scheme is consequently synchronizable.
\end{theorem}
\textit{Proof:}

-Desynchronization: Here we first highlight the fact that $A$ is not able to
produce valid messages for this protocol. Indeed, the only way to do this
is to know the secret key used either by the tag or the server. As the tag
is uncorrupted and the hash function is one-way, $A$ cannot learn anything
about this key. By blocking the last message of a protocol, $A$ desynchronizes
the tag as it updates its secret key contrary to the server. $A$ cannot use this
technique twice as the server resynchronizes its key during the search procedure in the second round. As a consequence, $Desync_S = 1$. The only way for $A$ to force the update of
$SK_i$ without updating $TK_i$ is to produce the first and second messages. As $A$ has
no information about $T_{max_i}$ and $TK_i$, he is not able to produce such messages. So,
$Desync_T = 0$. Finally, the scheme is ($1,0$)-desynchronizable.

-Resynchronization: By definition of the scheme, the tag is still accepted if
$TK_i$ is updated once. This is not the case if it is updated twice. So, $Resync_S = 1$.
If the server updates the stored key once, $TK_i$ is no longer stored in the
database and the server does not find a match, even if it updates all the
stored key once. As a consequence the tag is rejected, $Resync_T = 0$. The scheme
is ($1,0$)-resynchronizable and so synchronizable. $\square$

However, as discussed in \cite{AutoID}, eliminating any possibility of desynchronization is difficult on a technical level given the limited functionality of low-cost tags, and providing complete desynchronization resistant protocols is not the primary target of this papaer. Nevertheless, providing tools for detecting such an attack and localizing the adversarial device is not a major issue. In actual systems, the operator would have to physically remove or deactivate the attack device.

\noindent $\bullet$ \textit{Tag Inoperative}: In our schemes, the server can 
help an incapacitated tag to become operative by sending a renewed $T_{max_i}$ 
which is strictly greater than the previous one. This renewing capability also 
enables a server to willingly incapacitate a tag whenever it wants to, thus having 
full control over a tag. $\square$

\noindent $\bullet$ \textit{A note on Server Impersonation Attack}: Server impersonation \cite{impersonation:ubi} means that an adversary is able to impersonate a valid server
to a tag. One reason that this is a genuine threat is because desynchronization can
occur if a tag updates its stored data when the server does not. More specifically, an adversary that has
read a tag's stored secrets could impersonate an authorized server to the tag. If
the attacker executes an authentication session with the tag, impersonating a valid
server, then it could make the tag update its stored secrets, although the genuine
server will not update its stored data. The tag and the real server would then be
desynchronized, and incapable of successful communications. If an adversary (we consider active adversary here) has
access to all the exchanged messages and knows the tag's secrets ($k_i^j, T_{max_i}, T^{j-1}_{t_i}$) used in a
single authentication session, it can compute the refreshed $k_i^{j+1}$ for the following
session. Hence, our protocol only resists such an attack on the assumption
that an adversary does not have access to at least $T^j_{r_i}, R^j_{r_i}$ in an authentication session that is performed between an authorized
server and a tag, for which the adversary knows the tag's secrets ($k_i^j, T_{max_i}, T^{j-1}_{t_i}$). Such an assumption has been considered in \cite{impersonation:phd}.

Once an RFID tag's stored secrets have been compromised, it is difficult to prevent
server impersonation based desynchronisation attacks. Designing more robust RFID protocols that make such server
impersonation attacks more difficult to perform is out of the scope of our work. If an RFID protocol uses a digital
signature scheme for authentication of a server to a tag, then an active adversary is unable to
impersonate the server to a tag just by compromising the tag. However, the use of
public key cryptography may be beyond the capabilities of low-cost tags.

\section{Performance Analysis}
\subsection{Comparison with previous work}
We compare our schemes with previous works in this section. The comparison is mostly done with symmetric key-based 
protocols. However, we also pick up EC-RAC\cite{ecrac4} which is based on ECDLP. EC-RAC is shown to be achievable within around 14500
 gates which is lower than other public key-based authentication schemes. For clarity of comparison, we provide the same environment to the other works, i.e., 
aggregate function is also applied there, and DM-PRESENT-80 is assumed to be used in all the protocols as the 
required hash function. We assume that the hash 
functions have the same 
computational cost (time and resources) as the PRNG. From the tables we can see that not all protocols can satisfy 
the security requirements. Some of them also require high 
communication and computation cost. Moreover, reader authentication is 
also not supported by them. It is also important to keep the 
number of passes as low as possible. More than 2-pass protocols require 
more communication overhead, and a tag needs to `remember' all the 
intermediate states to complete the protocol. All this translates into needing 
more resources on the tag. Note that, our schemes provide mutual authentication through 2-pass. In protocols where the
 reader has to be authenticated first, the reader should broadcast its authentication value to the tags in the 
 environment. While this is also true for our schemes, the required computation (to authenticate a reader) by 
 a tag requires only one hash (as it has $T^{j-1}_i, T_{max_i}$ in its memory already).

Table 2 shows the comparison of security features like Reader Authentication, 
Forward Security, Backward Security, Replay, Cloning, Tracking, Timing Attack resistance, and whether compromise of a tag 
results in compromising other tags. For the sake of clear comparison, we apply the same assumptions on the other protocols as ours. Compared to all of the previous works, YA-TRAP* is the 
most efficient one. Unlike YA-TRAP*, our schemes do not allow 
a tag to become 
inoperative. Rather, the reader controls whether to disable a tag. Again, if one tag is compromised, the adversary should not be able to compromise the other tags based on the 
information gathered from the compromised tag. MSW protocol is vulnerable to such attacks. This concern does 
not arise in our protocols, since no two tags share any secrets between them. In 
other words, in our work, it is not possible for an adversary to derive secrets 
of other tags, even if he/she gets a tag's secrets. Moreover, tag tracking and cloning are the attacks that should be 
avoided to maintain a tag's privacy. Our protocols do not allow an adversary to track or clone a tag. 

\begin{table*}[hbp]
\caption{Performance Comparison (Security)} 
\begin{center}
\hbox to\hsize{\hfil
\begin{tabular}{|c|c|c|c|c|c|c|c|}

\hline
Features              & REP            & MSW               & EC-RAC\cite{ecrac4}         & YA-TRAP*      &  Scheme 1     & Scheme 2\\
\hline
Cloning Attack        & $\bigcirc$     & $\bigcirc$        & $\bigcirc$                  & $\bigcirc$    &  $\bigcirc$   &    $\bigcirc$\\
\hline
Timing Attack         & $\bigcirc$     & $\bigcirc$        & $-$                           & $-$             &  $\bigcirc$   &   $\bigcirc$\\
\hline 
Replay Attack         & $\bigcirc$     & $\bigcirc$        & $\bigcirc$                  & $\bigcirc$    &  $\bigcirc$   &   $\bigcirc$\\
\hline 
Tag Tracking          & $\bigcirc$     & $\bigcirc$        & $\bigcirc$                  & $\bigcirc$    &  $\bigcirc$   &   $\bigcirc$\\
\hline
Reader                & No             & No                & yes                         & No            & yes           &   yes\\
Authentication        & & & & & & \\  
\hline
Forward Secure        & yes            & No                & yes                         & yes           &   yes         &   yes\\
\hline 
Backward Secure       & yes            & yes               & yes                         & yes           &  yes          &   yes\\
\hline  
Other Tag             & $\bigcirc$     & $-$                 & $\bigcirc$                  & $\bigcirc$    & $\bigcirc$    &   $\bigcirc$\\
Compromise            & & & & & & \\
\hline
\end{tabular} \hfill}
\label{table cont.}
{\footnotesize {$\bigcirc$ = Not Vulnerable to Attack; $-$ = Vulnerable to Attack; Scheme 1 = Our original scheme; Scheme 2 = Extended version of Scheme 1 }}
\vspace{0.2cm} 
\caption{Performance Comparison (Cost)}
\hspace{-1.0cm}
\hbox to\hsize{\hfil
\begin {tabular}{|c|c|c|c|c|c|c|c|}

\hline
Features                      & REP            & MSW                       & EC-RAC\cite{ecrac4}       & YA-TRAP*     & Scheme 1  & Scheme 2  \\
\hline
Server Cost                   & O(n)           & O(nlog$_k$ n)             & O(n)                      & O(n)         & O(n)       & O(n)\\
\hline
Tag Comp                      & 1 Hash         & O(log n) Hash             & 5 scalar mul              & 5 Hash       &  4 Hash    &  5 Hash\\         
                              & &                                          & 2 Hash  & & &\\
\hline
Message flows                 & 2              & 2                         & 2                         & 2            &  2          &  2  \\
\hline
Tag Memory                    & 64 bit        & O(log n)$\times$ 64 bit    & 1913 bit                  & 193 bit      &  192 bit    &  192 bit  \\
\hline
T$\rightarrow$R Comm          & 2b             & 3b                        &  4b                       & 3b           &  2b         &  3b\\
\hline 
R$\rightarrow$S Comm          & 2nb            & 3nb                       &  4nb                      & (3n+2)b      &  (n+1)b     &  (n+1)b\\
\hline
R$\rightarrow$T Comm          & 3b             & 2b                        &  4b                        & 3b           &  3b        &  3b\\
\hline 
S$\rightarrow$R Comm          & 2b             & 2b                        &  4b                        & 3b           &  3b        &  4b\\
\hline
Gate Count                    & $\sim$ 2200    & $\sim$ 2200               & $\sim$ 14500               & $\sim$ 2200  & $\sim$ 2200 & $\sim$ 2200 \\
\hline
\end{tabular} \hfill}
\end{center}
\label{table cont.}

$\bullet${\footnotesize {all the protocols are assumed to utilize aggregate function; $n$ = total number of tags; $b$ = bit length of Messages (assuming all are equal 
in bit size); Memory requirement considers ROM and non-volatile RAM; Scheme 1 = Our original scheme; Scheme 2 = Extended version of Scheme 1 }}

\end{table*}

Table 3 shows the performance comparison based on cost. Considering that all the messages have same bit length $b$, 
the tag-to-reader message is 3$b$ bits long in Scheme 2 (2$b$ bits in Scheme 1), and the reader-to-server message is $(n+1)b$ 
bits long. Even if the previous works implement
aggregate function, our protocols achieve lower communication costs. As for the aggregate function, the extended version (Scheme 2) provides security through partial authentication, 
which is not used in our initial scheme (Scheme 1). We use partial 
authentication to keep rogue tags out of the aggregate function. This works as a filter to reject any possible fake 
tags. This partial authentication helps an aggregate function to be correctly 
verified by the server, hence authenticating the corresponding tags. The 
partial authentication requires a tag to compute one more hash function 
(computing $AT^j_{t_i}$) and contains $b$ more bits ($AT^j_{t_i}$) compared to Scheme 1 while 
communicating with the reader. We consider this to be a 
feature of Scheme 2; i.e., a performance trade-off for improved security. YA-TRAP* and our Scheme 2 require the same number of computations by a tag; i.e., 
4 hash functions and one PRNG, and Scheme 1 requires 3 hash functions and one PRNG. The bit length of reader-to-server communication is 
lower in both of our schemes than in YA-TRAP*. For $n$ number of tags, our schemes 
require $(n+1)b$ bits, whereas YA-TRAP* requires $(3n+2)b$ bits for the reader-to-server communication. In the initial
 phase, a server and a reader authenticate each other. We do not consider the server-to-reader communication cost of 
 the initial phase when they authenticate each other and when the server sends the required tag related information to the 
 reader. This communication can be done off-line. We are mainly concerned about the on-line communication costs. 
 However, reader-to-tag communication in our protocols are costlier than in MSW protocol. But our schemes
  satisfy more features than MSW scheme. Moreover, EC-RAC requires a processor for EC computation 
  and around 14500 gates which are well beyond today's low-cost tags. On the other hand, in REP, the protocol assumes an external device to be 
attached to the tag. This is not practical for applications where batch-mode 
authentication is done. About the memory requirement, our protocols require 192-bits of memory (64-bits for secret
 key $k_i^j$, 128-bits for timestamps $T^{j-1}_{t_i}$, $T_{max_i}$)- which is well within today's passive RFID 
 tags capability \cite{memory}.

\subsection{A quantitative performance analysis}
Before making our assumptions and analyzing our schemes, we briefly review some facts on practical deployment of an RFID system.

 \begin{itemize}
\item Tag readers are assumed to have a secure and dedicated connection to a back-end database. Although
readers in practice may only read tags from within the short tag operating
range, the reader-to-tag, or forward channel is assumed to be broadcast with a signal
strong enough to monitor from long-range. The tag-to-reader, or
backward channel is relatively much weaker, and may only be monitored by eavesdroppers
within the tag's shorter operating range \cite{Weis}. 

\item The time available for a complete reading/authentication procedure is in the range of 5-10 milliseconds considering the performance criteria of an RFID system that demands a minimum tag reading speed of at least 200 tags per second \cite{AutoID}.

\item In accordance with EPC C1G2 protocol, a maximum tag-to-reader data transmission rate bound of 640 kbps and a reader-to-tag data transmission rate bound of 126 kbps \cite{AutoID}. 

\item In the low-cost tags, the complexity of implementing
robust PRNGs is equivalent to the complexity of
implementing robust one-way hash functions. The same assumption has been widely considered in cryptographic literature, \cite{menezes,weim,Tsudik} to name a few.

\end{itemize}

$\textbf{Our assumptions:}$
Based on the above facts, we use the following assumptions for the quantitative analysis of our schemes.

\begin{itemize}
\item The tag reading speed is at least 200 tags per second.
\item The time for a complete reading/authentication procedure is in the range of 5-10 milliseconds.
\item The tag-to-reader data transmission rate bound is 640 kbps and a reader-to-tag data transmission rate bound is 126 kbps.
\item Computing a PRNG and a one-way hash function takes same time. 
\item DM-PRESENT-80 hash function is used as the underlying one-way hash function. It provides 64-bit security level, and operates in a single block with 33 cycles per block at the rate of 100 khz. Each of the hash function takes $\frac{33}{100 khz} = 0.33$ milliseconds to run on a tag. 
\item The time required for XOR and concatenation operations is ignored, since they take negligible amount of time and resource.
\item Computation time in reader and the server is ignored since reader and server have ample computational power. 
\end{itemize}

$\textbf{Quantitative performance:}$
 Our extended protocol requires 5 hash operations for a tag, thus taking $1.65$ milliseconds. The tag-to-reader communication requires 192-bits, so it will take $\frac{192 bits}{640 kbps}= 0.30$ milliseconds. Similarly, the reader-to-tag communication takes $1.52$ milliseconds. In total, our estimated total protocol execution time is: $(1.65 + 0.30 + 1.52) \approx 3.50$ milliseconds, which is well within the requirement as stated above. In the same way, the estimated time for Scheme 1 can be calculated. Assuming that 200 tags are read at one time, the estimated run times of our protocols are well within the bound. As for the required number of gates, our protocols are also well within the requirements for the low-cost tags, which are expected to have 2000-5000 gates available for security purposes \cite{AutoID}.

There might be some argument on the necessity of using aggregate functions. Most RFID readers have serial interfaces using RS/EIA 232 standards (point to point, twisted pair)\cite{essential}. Readers communicate with the back-end server using such an interface. RS 232 serial interface standard says that the bit rate is lower than 20,000 bits per second \cite{rs232}. As per our assumption, when 200 tags are read at a time in the batch-mode, it would take 25600-bits ($R_{t_n}$and $H_{id_n}$, where each of them are 64-bits long, and $n = 200$) to be transferred if no aggregate function is used, requiring around 1.28 sec to transfer the data. This complexity will grow linearly with the growth of the number of tags that are to be read in batch-mode. On the other hand, if we use aggregate function, it would take 12864-bits ($R_{t_n}$ and $H$, where aggregate value $H$ and each of $R_{t_n}$ are 64-bits long, and $n = 200$) to be transferred, thus requiring about 0.64 sec to transfer the data. In other words, the use of aggregate function will reduce the data transfer time by approximately 50$\%$. It is thus important to reduce the cost of reader-to-server communication, since 1 sec is quite a long time for cryptographic protocol execution online.

\section{Conclusion}

In this paper, we have proposed two low-cost and secure two-way authentication 
schemes without public key which are more efficient and secure than that in previous work. Our 
schemes are resistant to various attacks, and their required computation and 
communication costs are minimal. Furthermore, we show the use of aggregate hash 
functions in our schemes to reduce tag-to-reader and reader-to-server communication costs, which must be 
low for batch-mode 
authentication environment. In the one of our schemes, the reader 
uses partial authentication to keep rogue tags out of the aggregate 
function. This increases a tag's computation by one hash function, and also the 
tag-to-reader communication by $b$ bits. We consider this as a trade-off of efficiency for
security. Our protocols provide 
resistance against Cloning, Replay, 
Tracking, and Timing attacks. They also provide forward security, and do not allow a 
tag to become inoperative unless disabled by the reader. We have also shown that the performance of our protocols are in line with current RFID infrastructures. Such secure and efficient protocols are desired to realize 
intelligent and secure transportation system, supply-chain, inventory management, and many other applications fostering 
attractive information society

\label{sect:bib}
\bibliographystyle{plain}
\bibliography{easychair}

\end{document}